\documentclass[12pt]{article}
\usepackage{a4,amsmath,lscape,epsfig}
\def\ii{\'{\i}}
\def\beq{\begin{equation}}
\def\eeq{\end{equation}} 
\def\beqa{\begin{eqnarray}}
\def\eeqa{\end{eqnarray}}
\def\ban{\begin{eqnarray*}}
\def\ean{\end{eqnarray*}}
\def\bi{\begin{itemize}}
\def\ei{\end{itemize}}

\def\d{\mbox{d}}

\begin{document}

\begin{center}
{\bf Caloric curve for finite nuclei in relativistic models}
\end{center}

\begin{center}
{\it D.P.Menezes $^1$ and C. Provid\^encia $^2$}\\
{\it $^1$  Depto de F\'{\i}sica - CFM -
  Universidade Federal de Santa Catarina}\\
{\it Florian\'opolis - SC - CP. 476 - CEP 88.040 - 900 - Brazil}\\
{\it $^2$ Centro de F\ii sica Te\'orica - Depto de F\ii sica -
  Universidade de Coimbra}\\
{\it 3004-516, Coimbra - Portugal}\\
\end{center}

\vspace{0.50cm}

\begin{abstract}

In this work we calculate the
caloric curve (excitation energy per particle as a function
of temperature) for finite nuclei within the non--linear
Walecka model for different proton fractions. It is shown that the
caloric curve is sensitive to the proton fraction.
Freeze-out volume effects in the caloric curve are also studied.
\end{abstract}

\vspace{0.50cm}
PACS number(s): {\bf 21.10.-k, 21.30.-x, 21.65.+f,25.70.-z}
\vspace{0.50cm}

The production of several intermediate mass fragments in a short time scale
during heavy ion collisions is known as nuclear multifragmentation.
At several hundred MeV/u, the multifragment decay follows  the formation of an equilibrated projectile
remnant. The existence of equilibration  is consistent  with different
experimental observations, such as the
symmetry of the measured rapidity distributions of  fragments with
$Z\ge 3$ \cite{exp2}. The spectator matter has, therefore,
been used to investigate a thermally driven liquid-gas phase
transition \cite{exp2,exp1,exp3}. One of the evidences of this
transition in infinite systems is the fact that the heat--capacity
exhibits a peak at a given  temperature. However  in finite systems
the situation is more complicated \cite{exp1,Gross,INDRA,lbct}. The caloric
equation of state,  which is given by the
excitation energy per nucleon in terms of the thermodynamic temperature is
an important quantity to be investigated in the search for a signal of
a phase transition.  Nevertheless, it was recently pointed out that the
 identification of the existence of a phase transition cannot be
based only on the behaviour of the caloric curve and a more detailed
knowledge of the thermodynamic phase diagram is also required
\cite{Elliott}. In particular it was shown that the interpretation of the 
data is sensitive to the use of a variable free volume on the calculation.

Recently there has been a big development in the description of nuclei
and nuclear matter in terms of relativistic many-body theory. In
particular, the phenomenological models developed using the
relativistic mean-field theory describe well the ground-state of both
stable and unstable nuclei \cite{sed,toki94}. These same models are used to describe the
properties of neutron stars and super-novae \cite{toki98}. Therefore, it is
important to test these models  at finite temperature and different
densities. In particular, it would be interesting to compare the
caloric curve obtained within a relativistic Thomas-Fermi calculation
with the recent experimental data, and verify whether the
proton-neutron asymmetry of the hot source gives information on the
symmetry energy term of these models.

Within the framework of relativistic models, the liquid-gas phase transition
in nuclear matter has been investigated at zero and finite temperatures for
symmetric and asymmetric semi-infinite systems
\cite{toki98,rs,md,epsw,cev}.
With the help of the Thomas Fermi approximation, we have investigated
droplet formation in  the liquid-gas phase
transition in cold~\cite{mp1,nosso3} and hot~\cite{mp2} asymmetric nuclear
matters using  the non-linear Walecka model (NLWM)~\cite{sed,bb}.
As shown in Refs.~\cite{mp1,mp2}, the optimal
nuclear size of a droplet in a neutron gas is determined by a delicate balance
between nuclear Coulomb and surface energies. The surface energy favors nuclei
with a large number of nucleons $A$, while the nuclear Coulomb self-energy
favors small nuclei.

In the present work we calculate the caloric curve, given by the
temperature dependent excitation energy per particle, for the nuclei
obtained with the approach mentioned above. In particular, we will
study the influence of the proton-neutron asymmetry.  
We chose the systems $^{150}$Sm and  $^{166}$Sm because they lie in the
mass and charge  range of interest for the experiments we are analysing 
\cite{exp2,exp1,hagel}. In the first two references the
caloric curve presented is obtained with prefragments in the  mass
range 50 $\sim 100$ to 200. In the third reference the data were obtained 
for a compound nucleus of mass $\sim$ 160. We took two isotopes  with quite
different number of neutrons  in order to study the effect of
proton-neutron asymmetry. In the framework of the Thomas-Fermi theory, 
shell effects are washed out. Hence, we are
calculating average properties. We expect that the caloric curve of a
given system may  depend quantitatively on the system mass but the 
qualitative features, namely the dependence of proton-neutron
asymmetry and the effect of freeze-out volume, will be similar.   
In multifragmentation calculations an input parameter called the
freeze-out radius is normally used \cite{Gross}, so that a phase
transition at constant volume is simulated.  We  investigate
the consequences on the caloric curve when  thermalization in a
freeze-out volume is imposed in the present framework.

We start from the Lagrangian density of the
relativistic non-linear Walecka model \cite{bb}, \cite{ring}
$$
{\cal L}=\bar \psi\left[\gamma_\mu\left(i\partial^{\mu}-g_v V^{\mu}-
\frac{g_{\rho}}{2}  \vec{\tau} \cdot \vec{b}^\mu
-e A^\mu \frac {(1+ \tau_3)}{2}\right)
-(M-g_s \phi)\right]\psi
$$
$$
+\frac{1}{2}(\partial_{\mu}\phi\partial^{\mu}\phi
-m_s^2 \phi^2) - \frac{1}{3!}\kappa \phi^3 -\frac{1}{4!}\lambda
\phi^4
-\frac{1}{4}\Omega_{\mu\nu}\Omega^{\mu\nu}+\frac{1}{2}
m_v^2 V_{\mu}V^{\mu}
$$
\begin{equation}
-\frac{1}{4}\vec B_{\mu\nu}\cdot\vec B^{\mu\nu}+\frac{1}{2}
m_\rho^2 \vec b_{\mu}\cdot \vec b^{\mu}
-\frac{1}{4}F_{\mu\nu}F^{\mu\nu}\;,
\end{equation}
 where
$\phi$, $V^\mu$, $\vec{b^\mu}$ and $A^\mu$ are, respectively, the
scalar-isoscalar, vector-isoscalar, vector-isovector meson fields and the
electromagnetic field;
$\Omega_{\mu\nu}=\partial_{\mu}V_{\nu}-\partial_{\nu}V_{\mu}$ ,
$\vec B_{\mu\nu}=\partial_{\mu}\vec b_{\nu}-\partial_{\nu} \vec b_{\mu}
- g_\rho (\vec b_\mu \times \vec b_\nu)$
and
$F_{\mu\nu}=\partial_{\mu}A_{\nu}-\partial_{\nu}A_{\mu}$,
with the following parameters:
the nucleon mass $M=938$ MeV, the masses of
the mesons $m_s=492.25$ MeV, $m_v=795.36$ MeV, $m_{\rho}=763.0$ MeV, the
electromagnetic coupling constant $e=\sqrt{\frac{4 \pi}{137}}$ and
the self-interacting coupling constants $\kappa$ and $\lambda$.
The set of constants we use is normally identified as NL1
\cite{sed}, with $C_i^2= g_i^2 M^2/m_i^2$, $i=~s,~v,~\rho$, where
$C_s^2=373.176$, $C_v^2=245.458$, $C_{\rho}^2=149.67$,
$\kappa/M \times 10^{-3}=2 g_s^3 \times 2.4578$ and
$\lambda \times 10^{-3}=-6 g_s^4 \times 3.4334$. This parameterization
gives a good description of the ground-state properties of all stable
nuclei.

The thermodynamic potential is obtained  within the Thomas--Fermi 
approximation.  After it is minimized in terms of  the
meson and electromagnetic fields, the following coupled differential equations have to be solved:

\begin{equation}
\nabla^2 \phi=m_s^2\phi+
\frac{1}{2}\kappa \phi^2 +\frac{1}{3!} \lambda\phi^3 - g_s \rho_s,
\label{phi} \end{equation}

\begin{equation}
\nabla^2 V_0=
m_v^2 V_0 - g_v \rho_B
, \label{V0}\end{equation}
\begin{equation}
\nabla^2 b_0= m_\rho^2 b_0
-\frac{g_\rho}{2} \rho_3
, \label{b0}\end{equation}
\begin{equation}
\nabla^2 A_0=-e \rho_p
,\label{A0}\end{equation}
where
$$\rho_s= 2 \sum_{i=p,n}
\int \frac{\d^3p}{(2\pi)^3}
\frac{M^*}{\epsilon}\left(f_{i+}+f_{i-}\right),$$
with
\begin{equation}
f_{i\pm}({\mathbf r},{\mathbf p},t)\,=\,
\frac{1}{1+\exp[(\epsilon\mp\nu_i)/T]}\;
, \quad i=p,n,\end{equation}
where  $\nu_i\;=\mu_i-{\cal V}_{i0}$ are the effective chemical potentials
with $\mu_i$ being the chemical potentials for particles of type $i$ and
$$
{\cal V}_{p0}= g_v V_0  + \frac{g_\rho}{2} b_0 + e A_0\; ,
\quad {\cal V}_{n0}= g_v V_0  - \frac{g_\rho}{2}  b_0 \; ;
$$
$\epsilon=\sqrt{p^2+{M^*}^2}$,
$M^* =M-g_s\phi$ is the effective nucleon mass and $T$ is the temperature.
Moreover, $\rho_B=\rho_p+\rho_n,\, \rho_3=\rho_p-\rho_n$ with
\begin{equation}
\rho_i=2 \int\frac{\d^3p}{(2\pi)^3}(f_{i+}-f_{i-}), \quad i=p,n\; ,
\label{rhoi}\end{equation}
and the energy density, obtained from the thermodynamic potential reads:
$${\cal E}(r)=\,2 \sum_i \int\frac{d^3p}{(2\pi)^3}\,
\left[\epsilon(f_{i+}+f_{i-}) +{\cal V}_{i0}(f_{i+}-f_{i-})
\right]$$
$$+\frac{1}{2}
\left[(\nabla \phi)^2 -(\nabla V_0)^2 - (\nabla b_0)^2 - (\nabla A_0)^2
\right] $$
\beq
+\frac{1}{2} \left(
m_s^2 \phi^2 + \frac{2}{3!} \kappa \phi^3 + \frac{2}{4!} \lambda \phi^4
-m_v^2 V_0^2  -m_\rho^2 b_0^2  \right). \label{dens}
\end{equation}

The coupled differential equations are solved numerically. For more details
on the analytical and numerical procedure, please refer to \cite{mp1,mp2}.
Three kinds of instabilities can occur in this system.
The condition for mechanical stability requires that
$ \left(\frac{\partial P}{\partial\rho_B}\right)_{Y_p}\ge 0\;$,
where $P$ is the pressure and $Y_p=\rho_p/\rho_B$ is the proton fraction.
The condition for diffusive stability implies the
inequalities
$ \left( \frac{\partial \mu_p}{\partial Y_p} \right)_{P,T} \ge 0$
and
$ \left( \frac{\partial \mu_n}{\partial Y_p} \right)_{ P,T} \le 0\;$.
Finally, the thermodynamical stability is expressed by
$C_v=\left(\frac {d\varepsilon^*}{dT}\right)_{v, Y_p} > 0$,
where $C_v$ is the specific heat and
$\varepsilon^*=\varepsilon(T)-\varepsilon(T=0)$ is the excitation energy per
particle, with
the total energy per particle at any temperature given by \cite{shlomo1}
$ \varepsilon(T)=\int \frac{{\cal E}(r)}{A} d^3r $, where $A= Z+N$.
The two-phase liquid-gas coexistence is governed by the Gibbs condition.

We have first solved the equations of motion for an infinite system
in order to obtain appropriate boundary conditions
for the program which integrates the set of
coupled non-linear differential equations (\ref{phi}) to (\ref{A0})
in the Thomas-Fermi approximation. Once the fields are obtained,
all thermodynamic quantities of interest can be easily calculated.
The binding energy per nucleon is { $\frac{B}{A}=\varepsilon(T)-938.$
MeV.

In table 1, we show the binding energy per nucleon
and the excitation energy per particle for the $^{150}_{62}$Sm$_{88}$,
which has a proton fraction equal to $0.41$. In table 2, the same quantities
are shown for the  $^{166}_{62}$Sm$_{104}$, with a proton
fraction of $0.37$. Notice that, independently of the proton fraction,
the excitation energy per particle increases with  temperature in the range of temperatures shown.  For higher
temperatures  we were not able to obtain
convergence for a droplet of the size considered.

In tables 3 and 4, we give  the binding and
the excitation energies per particle when a freeze-out volume of respectively
$6V_0$ and $9V_0$ is used  with $V_0$ the volume at
$T=0$. We have considered a freeze-out radius of $2.2 A^{1/3}$ fm for
$6V_0$ and $2.5 A^{1/3}$ fm for  $9V_0$ with $A=166$. In this case the
solutions obtained consist of  a droplet immersed in a gas of
evaporated particles, in such a way that they mimic a source of changing
mass. As temperature increases more particles
evaporate, mainly neutrons,  and the fraction of protons in the
droplet increases. This can be seen in  tables 3 and 4  where the
number of particles which remain inside the droplet as well as  the
droplet proton fraction ($Y_d$) are given. We conclude that the larger the
freeze-out volume the faster the excitation energy increases with temperature
and the larger is the proton fraction in the droplet.  This picture
is consistent with the discussion presented in \cite{lbct}.

The results for the excitation energies shown in all tables are displayed
in figure 1. Also shown are the experimental data of
refs. \cite{exp2,exp1}, and the Fermi-gas law $\varepsilon^*= 1/k\,T^2$, with
$k=10.0$ (thin dashed line) and 13.0 (thin full line).  We have considered
that the measured temperature $T_{\mbox{\small HeLi}}$
($T_{\mbox{\small HeTD}}$),
obtained from the isotope yield ratios $^3$He/$^4$He and $^6$Li/$^7$Li
($^3$He/$^4$He and $^2$H/$^3$H), satisfy, in the range of densities
considered,  $T_{exp}/T\sim 0.85$ and have scaled the experimental
data accordingly \cite{exp1}.

We conclude that the excitation energy for $^{166}$Sm (thick long-dashed curve), proton
fraction $0.37$, increases slightly slowlier with temperature than for $^{150}$Sm (thick full
curve), proton fraction $0.41$,  although the difference is not large. These two curves
are consistent with data of \cite{hagel} and a level density parameter
$A/k,\, k=13.0$ in the Fermi gas  model relation.
This  agrees with the observed value at around  2 MeV excitation
energies \cite{hagel}.
 Experimental results obtained at higher bombarding energies
\cite{exp2,exp1} give higher excitation energies for the same
temperatures. It can be seen from figure 1 of ref. \cite{exp1} that the
higher excitation energies correspond to smaller sources. The larger
sources with an average $Z_{\mbox{bound}} \ge 60$ have
excitation energies  $\varepsilon^*\le 5$ MeV. In the present approach
the solutions obtained in a fixed volume, correspond to droplets in  a gas of
free particles. These solutions have higher excitation energies than
the ones obtained with no { \em a priori} fixed volume. In average,  this situation
corresponds to smaller systems at higher energies. This could explain the
change of slope that is observed in the calculated data  both for
$V=6 \, V_0$ and $V=9 \, V_0$, in such a way that they come closer to the
experimental data \cite{exp2,exp1}. The same effect was obtained in
ref. \cite{bugaev2}, where an exact analytical solution of the statistical 
multifragmentation model was found in the thermodynamic limit. 
For a fixed nucleon density, the caloric curve rises more slowly for 
lower densities and its leveling occurs at lower temperatures. The leveling 
of the caloric curve is associated with the fast 
change of the configurations from a state dominated by one liquid fragment to 
a gaseous multifragment configuration. We can draw a similar conclusion from 
tables 3, 4 and figure 1: the leveling of the thick
dash--dotted and short-dashed curves occurs faster when the droplet (liquid 
phase) becomes smaller. It would be interesting to study the effect of the 
symmetry energy on the leveling of the caloric curve in the 
statistical multifragmentation model of ref. \cite{bugaev2}.

In summary, we have studied the excitation energies
of arising droplets in a vapor system, up to  $T=6.5$ MeV. The
droplets are described in terms of a non-linear Walecka--type model within the
Thomas--Fermi approximation.
We have used the NL1 parameterization, which is known to describe well
the ground-state properties of nuclei.  The excitation energies of droplets either corresponding to
$^{150}$Sm or  $^{166}$Sm, for temperatures between 3 and 6.5 MeV,
 are consistent with  the caloric curve in  the Fermi gas approximation with
 a level density parameter  A/13. This result agrees with experimental
 data obtained in heavy-ion collisions at intermediate energies
 \cite{hagel}. We have shown that the caloric curve is
 sensitive to the proton fraction  and therefore to the symmetry term
 of the model used.  Experimentally the dependence on the proton
 fraction could be studied by comparing data obtained
 from sources with different proton fractions.
 For the range of temperatures
 studied,
the NL1 parameterization of the non-linear
 Walecka model has shown to be adequate to describe nuclear properties
 and therefore it is a good candidate to generate an equation of
state  for astrophysical purposes.

When a freeze-out radius is imposed, our procedure yields
caloric curves which come closer to the experimental results obtained in
heavy -ion collisions at higher energies \cite{exp2,exp1}. In this
case we have smaller droplets with a higher proton fraction immersed in
a gas of particles, mainly neutrons. This could be interpreted as an
oversimplified picture
of the second regime in the statistical model prediction
\cite{bondorf}, namely the coexistence phase with a multifragment mixture.
This interpretation is supported by the results of ref.
\cite{bugaev2}.

 Although  the thermodynamical
equilibrium analysis oversimplifies the problem of high energy
heavy-ion collisions it is useful for providing a concrete description
of  warm nuclei and for showing qualitative features that should
be present in more microscopic calculations.

\vskip 0.35in
\begin{center}
{\bf Acknowledgments}
\end{center}

We acknowledge the computation facilities offered by Centro de F\'{\i}sica
Computacional of the University of Coimbra. C. P. would also like to
thank the warm hospitality in the Departamento de F\ii sica, Universidade
Federal de Santa Catarina. This work was partially supported by CNPq - Brazil
and
CFT - Portugal under the contracts PRAXIS/2/2.1/FIS/451/94 and
POCTI/1999/FIS/35308.

\newpage

\begin{table}
{\bf Table 1.}
Output results given by the solution of the coupled differential
equations for different temperatures for $^{150}_{62}$Sm$_{88}$ ($Y_p=0.41$).
\vspace{0.5cm}
\begin{center}
\begin{tabular}{ccc}
\hline
$T$ & $B/A$  & $\varepsilon^*(T)$
\\
(MeV) & (MeV/A) & (MeV/A)
\\
\hline
\hline
0.   & -8.2 & 0.0 \\
3.   & -7.6 & 0.7  \\
4.   & -7.1 & 1.2  \\
5.  & -6.4 & 1.8 \\
6.  & -5.4 & 2.8 \\
6.5 & -4.8 & 3.4 \\
\hline
\end{tabular}
\end{center}
\end{table}

\vspace{0.5cm}

\begin{table}
{\bf Table 2.}
Output results given by the solution of the coupled differential
equations for different temperatures for $^{166}_{62}$Sm$_{104}$ ($Y_p=0.37$).

\begin{center}
\begin{tabular}{ccc}
\hline
$T$ & $B/A$  & $\varepsilon^*(T)$
\\
(MeV) & (MeV/A) & (MeV/A)
\\
\hline
\hline
0    & -7.9 & 0.0 \\
2.   & -7.7 & 0.3 \\
2.5  & -7.5 & 0.4\\
3.   & -7.3 & 0.7\\
4.   & -6.8 & 1.1\\
5.   & -6.1 & 1.8\\
6.   & -5.2 & 2.7\\
6.5  & -4.8 & 3.1 \\
\hline
\end{tabular}
\end{center}
\end{table}

\vspace{0.5cm}

\begin{table}
{\bf Table 3.}
Output results given by the solution of the coupled differential
equations for different temperatures for $^{166}_{62}$Sm$_{104}$
in a fixed volume, $6V_0$.

\begin{center}
\begin{tabular}{cccccc}
\hline
$T$ & $B/A$& $\varepsilon^*(T)$ & $A$ & $Y_d$\\
(MeV)&(MeV/A)&(MeV/A)&&\\
\hline
\hline
0.  & -7.9 & 0.0 & 166 & 0.37\\
3.  & -7.1 & 0.8& 163 & 0.38\\
4.  & -6.3 & 1.6& 158 & 0.39\\
5.  & -5.4 & 2.5& 154 & 0.40\\
6. & -3.9 & 4.0& 148 & 0.40\\
6.5& -3.0 & 4.9& 143 & 0.40\\
\hline
\end{tabular}
\end{center}
\end{table}

\vspace{0.5cm}

\begin{table}
{\bf Table 4.}
The same as figure 3 for a fixed volume, $9 V_0$.

\begin{center}
\begin{tabular}{cccccc}
\hline
$T$ & $B/A$& $\varepsilon^*(T)$ & $A$ & $Y_d$\\
(MeV)&(MeV/A)&(MeV/A)&&\\
\hline
\hline
0.  & -7.9& 0.0 & 166 & 0.37\\
3.  & -7.0& 0.9& 161 & 0.38\\
4.  & -6.2& 1.7& 156 & 0.39\\
5.  & -5.1& 2.8& 150 & 0.41\\
6.  & -3.4& 4.5& 142 & 0.41\\
6.5& -2.2& 5.3& 134 &0.41\\
\hline
\end{tabular}
\end{center}
\end{table}

\begin{figure}
\begin{center}
\epsfig{file=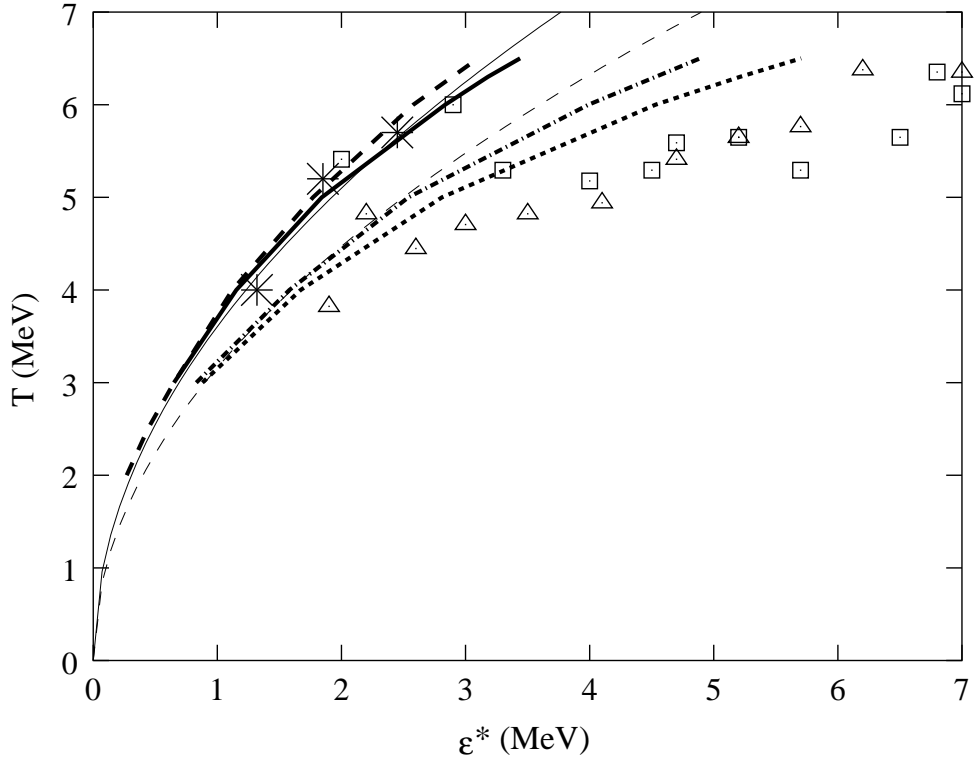,height=10cm,angle=0}
\caption{The caloric curves are shown for
$^{166}_{62}$Sm$_{104}$ ($Y_p=0.37$ - thick long-dashed  line),
$^{150}_{62}$Sm$_{88}$ ($Y_p=0.41$ - thick full line);  at $6\,V_0$
(thick dash-dotted line)
and $9\,V_0$ (thick short-dashed line) fixed volumes for $^{166}$Sm,  and
for the  Fermi gas law  \cite{hagel} ($k=10.0$ - thin dashed line and
$k=13.0$ - thin full line). Experimental results from \cite{exp1} (squares),
\cite{exp2} (triangles) and \cite{hagel} (big stars) are also displayed.}
\label{fig1}
\end{center}
\end{figure}

\end{document}